\documentclass[twocolumn, showpacs,preprintnumbers]{revtex4}
\usepackage{amsmath,amssymb}
\usepackage{graphicx}
\usepackage{dcolumn}
\usepackage{bm}
\usepackage{color}

\def \dd#1{\null}

\begin{document}

\title{Deterministic production of free-propagating Fock states of programmable photon number from a single atom-cavity system}
\author{A. Gogyan,$^{1,2}$ S. Gu\'erin,$^{2}$ C. Leroy,$^{2}$ and Yu.
Malakyan$^{1,3}$}
\email{yumal@ipr.sci.am}
\affiliation{$^{1}$ Institute for Physical Research, Armenian National Academy
of Sciences, Ashtarak-2,
0203, Armenia}
\affiliation{$^{2}$Laboratoire Interdisciplinaire Carnot de Bourgogne, UMR CNRS
6303, BP 47870, 21078 Dijon, France}
\affiliation{$^{3}$ Centre of Strong Field Physics, Yerevan State University, 1
A. Manukian St., Yerevan 0025, Armenia}

\date{\today }

\begin{abstract}

We propose a mechanism for producing Fock states on demand leaking from a single
mode optical cavity interacting with a single atom and a laser pulse. The number of photons can be chosen, as it is determined by the Zeeman substructure
of the ground state of the atom and its initial state. The deterministic generation of a free-propagating Fock state of $1\leq n\leq2F$
photons is achieved, when a circularly polarized laser pulse completely transfers the atomic population
between Zeeman sublevels of the ground hyperfine state $F$ through far-detuned Raman scattering thus
producing linearly polarized cavity photons. We describe analytically the evolution of optical field taking
into account the spontaneous losses and the cavity damping. We demonstrate the possibility of production of Fock-state
with different numbers of photons by using different transitions of the same atom. We show
also that this technique provides a deterministic source of a train of identical multiphoton Fock-states, if a
sequence of left- and right-circularly polarized laser pulses is applied. The resulting states have potential
applications in quantum computation and simulation.

\end{abstract}

\pacs{32.80.Qk %Coherent control of atomic interactions with photons
42.50.Ex; %Optical implementations of quantum information processing and transfer
03.67.Hk; %Quantum communication
42.50.Pq %Cavity quantum electrodynamics; micromasers
} \maketitle

% Force line breaks with \\

% It is always \today, today,
%  but any date may be explicitly specified

% PACS, the Physics and Astronomy
% Classification Scheme.
%\keywords{Suggested keywords}%Use showkeys class option if keyword
%display desired

\section{\protect\normalsize INTRODUCTION}

The deterministic production of multi-photon Fock-states (FS) is one of the essential
issues in contemporary quantum information technology, especially for the physical implementation
of secure quantum communication \cite{1,2}, linear optical quantum computing \cite{3}, high-precision
quantum interferometry \cite{4}, engineering of novel types of strongly correlated quantum systems
\cite{5} etc. For this purpose many schemes have been proposed so far based on the key idea to
achieve a strong coupling of atoms and photons for robust and efficient control of photon
production. One major trend is to use various designs of atom-high-finesse optical cavity
systems, where the frequency, polarization and propagation direction of the photons are given
by the atom-cavity properties, while the temporal profile of the photon distribution can be
controlled by the laser pulse interacting with the atom. Most of the studies have considered
so far a sequence generation of single-photons \cite{6}. Meanwhile, the creation of complex quantum
states of traveling light such as coherent superposition of photon-number states requires multi-photon
Fock states with different numbers of photons \cite{7}. With a train of atoms, the possibility
of producing intracavity multi-photon FSs was explored in the schemes based on
adiabatic transfer of atomic ground-state Zeeman coherences in single atoms \cite{8} or on vacuum
Rabi oscillations with long lived excited atomic state \cite{9} provided the atomic velocity is
conveniently tuned. In both cases, the photons generated by individual atoms remain
stored in the cavity due to negligibly small cavity decay rate. A scheme for the deterministic
generation of $N$-photon FS from $N$ three-level atoms confined in a high-Q cavity has
been proposed in \cite{10}. In all these processes, at least one atom is required for one photon
production. In contrast, the generation of two-photon FS by a single atom
coupled to a high-finesse cavity possessing, however, two non-degenerate modes has
been reported in \cite{11}. Recently, a non-linear mechanism allowing the production of multi-photon FS by a single mode cavity and a bichromatic pulsed field interacting with a single two-state atom has been proposed \cite{Bichrom}. However, this process is relatively sensitive to spontaneous emission.

\begin{figure*}[t] \rotatebox{0}{\includegraphics*
[scale =0.7]{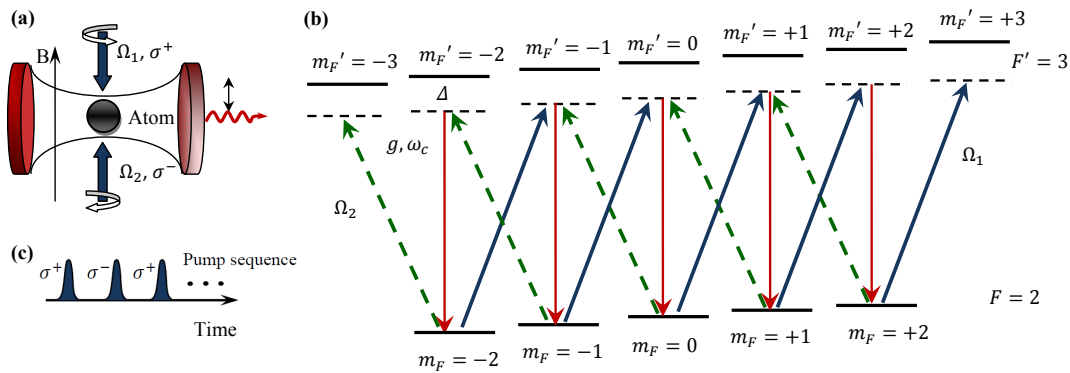}} \caption{(Color online) (a) Schematic setup. A single atom trapped in a high-Q cavity is driven by two
laser pulses. (b) $5S_{1/2}(F = 2) \leftrightarrow 5P_{3/2}(F' = 3)$ transition level structure of $^{87}$Rb in an external magnetic
field. $\Omega_1 (\Omega_2)$ laser is shown by thick blue solid (green dashed) lines and the red thin solid lines show the cavity-mode coupling. (c) Sequence of laser pulses.\label{Fig1}}
\end{figure*}

In this paper, we propose a deterministic source of multi-photon FS consisting of
a single $F$-Zeeman-structured multi-level atom, with $-F\le m_F\le +F$,  strongly coupled to a single-mode optical cavity. In
our scheme, an alkali atom prepared in one of the ground Zeeman sublevels, labeled $m_{F_i}$, interacts with a
circularly $\sigma_+$ polarized laser field and a linearly polarized cavity mode in a far-resonant Raman
configuration that results in production of cavity photons by transferring the atom into the Zeeman sublevels $m_{F}>m_{F_i}$ (see Fig. 1). The process stops, when the atomic population is completely transferred into the extreme Zeeman state $m_F=+F$. The emitted photons leave the cavity through one mirror as a freely
propagating wave packet, whose temporal profile is determined by the shape of laser pulse. A Fock state comprising $2F$ photons can be emitted at maximum, when the initial state is $m_{F_i}=-F$.

In
the context of diverse quantum information applications, this is an important advance
compared to the case of intracavity generation of multi-photon FSs \cite{8}, where the
Zeeman substructure of atomic ground state was similarly used. The off-resonant Raman
interaction makes the system immune to spontaneous losses into field modes other than the
cavity mode, thus ensuring the robust and efficient production of Fock-states, while the number
of photons depends on the initial state preparation and is limited rather by the number of Zeeman
sublevels than the laser field parameters.

One can produce a train of identical multi-photon FSs comprising $2F$ photons using a sequence of alternating left- and right-circularly polarized laser pulses.
After passage of the first laser pulse, a second pulse of
the same frequency, but with the orthogonal polarization drives indeed the atomic population
from $m_F=+F$ to $m_F=-F$ through consecutive Zeeman sublevels by emitting the same Fock state comprising $2F$ photons. This scheme provides a generalization of a technique that has been developed in the work \cite{SP}
for producing a train of indistinguishable single-photons on demand from an atom confined in a
high-Q cavity. The main advantage of this approach is that, in contrast to previous schemes \cite{6},
the cavity photons are produced without repumping the atom between photon generations.

This paper is organized as follows. In the next section we present the interaction setup
and atomic system and derive the basic equations for the time evolution of the atomic state
amplitudes and the cavity field. Here we estimate the error due to spontaneous losses and discuss
the main approximations ensuring the deterministic generation of Fock-states. We also find the
analytic solutions for the flux and numbers of output photons. In Sec. III we analyze via
numerical calculations the dynamics of atomic population and temporal distribution of output
photons. Our conclusions are summarized in Sec. IV.

\section{\protect\normalsize MODEL AND BASIC EQUATIONS}

\subsection{The Hamiltonian}
 The present mechanism for producing multi-photon Fock-state is based on the recently proposed method of deterministic generation of a stream of single-photon pulses in a single-atom
- single-mode cavity QED system \cite{SP}. A multi-level atom or ion is trapped in a one-mode high-finesse cavity and interacts with a $\sigma^+$-polarized laser field
$\Omega_1$ (Fig. \ref{Fig1}a) on the multi-level chain, for instance, on the cycling transition $5S_{1/2}(F = 2) \rightarrow 5P_{3/2}(F' = 3)$ of a $^{87}$Rb atom (Fig. \ref{Fig1}b), where the state
$5P_{3/2}(F' = 3)$ is well isolated from other hyperfine levels. For the production of a train of Fock-states, we apply a second laser pulse $\Omega_2$ with
orthogonal polarization $\sigma^-$, which acts on the atom after $\sigma^+$-pulse with a programmable delay time, larger than the laser pulse duration (Fig. \ref{Fig1}c) (see Sec. III ).

The atomic states are split into Zeeman sublevels by an external magnetic field perpendicular to the propagation axis of the laser fields in order to prevent the
mixing of ground Zeeman-states population by an ambient magnetic field. A coherent $\sigma^+$ polarized field of Rabi frequency $\Omega_1$ coupling the ground
state $F=2$ with magnetic quantum number $m_F = -F,\cdots,F$ and the excited state $|F' =3, m_{F'} = m_F+1\rangle$ creates a linearly polarized cavity-mode Stokes-photon (shown in
Fig.\ref{Fig1}b by red thin lines) via vacuum-stimulated Raman process on the transition $|F' =3, m_{F'} = m_F+1\rangle \rightarrow |F =2, m_F+1\rangle$, and thus
transfers the atom into the next Zeeman sublevel with $m_F+1$.
%For further convenience, we suppose that the atom-field coupling constants $g_{FF'}$ on $|F', m_{F'}\rangle
%\rightarrow |F, m_F\rangle$ transitions do not depend on the magnetic quantum numbers.
As a simplification in the notations, we consider that the atom-field coupling constants on $|F', m_{F'}\rangle
\rightarrow |F, m_F+i\rangle$ transitions do not depend on the magnetic quantum numbers. This actual dependence is considered in the numerical simulations.

The laser fields are tuned to the two-photon resonances, while the one-photon detuning $\Delta$ is very large compared to the cavity damping rate $k$, the spontaneous decay rate $\gamma_{\text{sp}}$ of the atom and the Rabi and Larmor frequencies: $\Delta \gg k, \gamma_{\text{sp}}, \Omega_{1,2}, \Delta_B^{(F, F')}$, where $\Delta_B ^{(F,F')}=g_L ^{(F,F')}\mu_B B$ is the Zeeman splitting of the ground and excited states in the magnetic field $B$, with $g_L^{(F, F')}$ the Land\'e factor and $\mu_B$ the Bohr magneton. This condition allows one to neglect the spontaneous losses from upper levels and dephasing effects induced by other excited states. In the far off-resonant case and for slowly varying laser fields: $d\Omega_i / dt \ll \Delta\Omega_i$, one can adiabatically eliminate the upper atomic states that leads to the effective Raman atom-photon coupling
\begin{equation} \label{defG}
G_i=g\Omega_i/\Delta, i=1,2,
\end{equation}
which can be made much slower than the cavity field decay: $G_i\ll k$. This ensures that the generated photons leave the cavity without being reabsorbed by the atom, resulting in a deterministic emission of photons.

We describe the dynamics of the system in the Heisenberg picture. The pumping laser fields
propagate perpendicular to the cavity axis and are given by
\begin{equation}\label{1}
E_j(t)={\mathcal E}_jf_j^{1/2}(t)\exp (-i\omega _jt),\ j=1,2,
\end{equation}
where $f_j(t)$, $j=1,2$, features their temporal profile of duration
$T$ and ${\mathcal E}_j$ is the peak amplitude of the field $j$ at frequency
$\omega_j$.

In this section we consider the case, when only the $\sigma^+-$pump pulse is applied. The effective interaction Hamiltonian in the RWA takes the form
\begin{equation}\label{hamiltonian}
H=\hbar \left[G_1 f_1^{1/2}(t)\sum_{m_F=-F}^{F-1}\sigma_{m_F,m_F+1}(t)a^{\dag}(t)+h.c.\right]
\end{equation}
with $\sigma_{ij}(t)$ and $a(t)$, $a^\dag(t)$ the atomic and cavity mode operators, respectively, in the Heisenberg representation. The peak Rabi frequencies of the
laser fields are given by $\Omega _{1}=\mu_{J,J'}\mathcal E_{1}/\hbar$ with $\mu _{J,J'}=\langle J||D||J'\rangle$ the dipole matrix element of the $J\leftrightarrow J'$ transition and $D$ the dipole moment operator.
We have omitted the Stark shifts of the atomic ground states induced by $\Omega_1$ laser pulse and cavity field as they are negligibly small as compared to the cavity decay rate $k$.

%of the form and $g^2/\Delta$ are negligibly small with respect
%to the cavity linewidth $k$ in the limit $G_1\ll k$, as considered here.

\subsection{Equations for the atomic populations}

The equations for the Zeeman sublevel populations $\langle\sigma_{m_F}(t)\rangle$ and ground-state coherences
$\langle \sigma_{m_F,m_{F}+1} (t) \rangle$ are derived from
the master equation for the whole density matrix $\rho$ of the system in the limit $G_1\ll k$ \cite{SP}. All two-photon processes $|F,m_F\rangle \rightarrow |F,m_F+1\rangle$ corresponding to an absorbtion of one laser photon from the $\sigma^+$ pump pulse and an emission of one cavity photon are taken having the same probability according to the present approximation of equality of the optical transition dipole moments. We obtain
%\begin{widetext}
%\begin{subequations}
\begin{eqnarray}\label{mFock}
{d\langle \sigma_{m_F} (t) \rangle\over dt}&=&A^{(1)}_{m_F-1}(t)\langle \sigma_{m_F-1}(t) \rangle \nonumber \\
&&- [A^{(1)}_{m_F}(t)+\Gamma^{(1)}_{m_F,m_F+2}(t)]\langle \sigma_{m_F}(t) \rangle \nonumber \\ &&+\Gamma^{(1)}_{m_F-2,m_F}(t)\langle \sigma_{m_F-2}(t) \rangle
\end{eqnarray}
\begin{equation}\label{diags}
{d\langle \sigma_{m_F,m_{F}+1} (t) \rangle \over dt} = - \frac{1}{2}[\alpha_1f_1(t)+\Gamma_{1}(t)]\langle \sigma_{m_F,m_{F}+1}(t)\rangle.
\end{equation}
with $A^{(1)}_{m_F<-F}=A^{(1)}_{m_F\ge F}=0$ and $\Gamma^{(1)}_{m_F,m_F+i}=0$ for $m_F+i > F$ or $m_F<-F, i=1,2$.
%\end{subequations}
%\end{widetext}
The rates are denoted
\begin{eqnarray}
A^{(1)}_{m_F}(t)&=& f_1(t)\alpha_{1}+\Gamma^{(1)}_{m_F,m_F+1}(t),\\
\Gamma_{1}(t)& =& {\Omega_1^2 \over \Delta^2}f_1(t)\gamma_{\text{sp}},\\
\Gamma_{m_F,m_{F}+i}^{(1)}(t)& =& {\Omega_1^2 \over \Delta^2}f_1(t)\gamma_{F',m_{F'},F,m_F+i},\qquad \\
& & m_{F'}=m_F+1,\quad i=1,2, \nonumber \\
\alpha_1 &=&4G_1^2/k,
\end{eqnarray}
where the partial decay rate $\gamma_{F',m_{F'},F,m_F+i}, \ i=1,2$ of state $| F',m_{F'} \rangle$ is of the same order as the spontaneous decay rate $\gamma_{\text{sp}}$. Correspondingly, $\Gamma_{m_F,m_{F}+i}^{(1)}(t)$ features the optical pumping (OP) from $|F,m_F\rangle$ into states $|F,m_{F}+i\rangle, i = 1,2$, induced by the $\Omega_1$ pump pulse. One can easily check from \eqref{mFock} that the total population of atomic ground state is conserved:
\begin{equation}\label{sum}
\sum_{m_F=-F}^F\langle \sigma_{m_F} (t)\rangle  = 1.
\end{equation}
Equation \eqref{diags} is solved with the initial condition $\langle \sigma_{m_F, m_{F}+1}(-\infty)\rangle =0$, so that the ground-state
coherence remains zero at all times.
It is seen from Eq. \eqref{mFock} that state $|F,m_F\rangle$ is populated in two ways: (i) via cavity photon generation with rate $\alpha_{1}f_1(t)$ and by the OP of rate $\Gamma_{m_{F}-1,m_F}^{(1)}(t)$ both from state $|F,m_{F}-1\rangle$ described by the first term in \eqref{mFock} and (ii) by optical pumping of rate $\Gamma_{m_{F}-2,m_F}^{(1)}(t)$ from state $|F,m_{F}-2\rangle$. At the same time it is depleted via OP into states $|F,m_{F}+1\rangle$ and $|F,m_{F}+2\rangle$ given
by the second term in \eqref{mFock}.

\subsection{Connection of the outgoing photon flux to the atomic populations}

The flux of the outgoing photons is defined by
\begin{equation}\label{flux}
\frac{dn_{\text{out}}}{dt}(t) = \langle a^{\dag}_{\text{out}}(t)a_{\text{out}}(t)\rangle,
\end{equation}
which determines the the shape of emitted field. Here $n_{\text{out}}(t)$ is the mean photon number of the output field $a_{\text{out}}(t)$ from the cavity which is connected to the input $a_{\text{in}}(t)$ and intracavity $a(t)$ field operators by the input-output relation \cite{gard}
\begin{equation}\label{7}
a_{\text{out}}(t)-a_{\text{in}}(t)=\sqrt{k}a(t)
\end{equation}
and satisfies the commutation relation $[a_{\text{out}}(t),a_{\text{out}}^{\dag}(t^{\prime})]=[a_{\text{in}}(t),a_{\text{in}}^{\dag}(t^{\prime})]=\delta (t-t^{\prime })$.
The operators $a(t), a^\dagger(t)$ are
obtained from the Heisenberg-Langevin equation \cite{gard} along with the
Hamiltonian \eqref{hamiltonian} as
\begin{align}
\dot{a}&=-iG_1 f_1^{1/2}(t)\sum_{m_F=-F}^{F-1}\sigma_{m_F,m_F+1}(t)\nonumber\\
&\qquad-{k\over 2} a(t)-\sqrt{k}a_{\text{in}}(t).
\end{align}

For $kT\gg 1$, we adiabatically eliminate the cavity mode $a(t)$, taking $\dot a=0$, yielding
\begin{equation}
a(t)=-\frac{2i}{k}G_1 f_1^{1/2}(t)\sum_{m_F=-F}^{F-1}\sigma_{m_F,m_F+1}(t) - \frac{2}{\sqrt{k}}a_{\text{in}}(t). \label{five}
\end{equation}
Upon substituting \eqref{five} into the flux equation \eqref{flux} and using \eqref{7}, for the vacuum input $\langle a^{\dag}_{\text{in}}(t) a_{\text{in}}(t)\rangle = 0$,
we obtain
\begin{equation}\label{nout}
\frac{dn_{\text{out}}}{dt}(t)=\alpha_1f_1(t)\sum_{m_F=-F}^{F-1}\langle \sigma_{m_F}(t)\rangle,
\end{equation}
describing the generation of cavity photons by the $\Omega_1$ field initialized from the Zeeman sublevels $-F \leq m_F \leq F-1$.

Upon substitution of \eqref{sum} into Eq. \eqref{nout}, one finds
\begin{equation}\label{nout15}
\frac{dn_{\text{out}}(t)}{dt}=\alpha_1f_1(t)[1-\langle \sigma_{F}(t)\rangle].
\end{equation}
We see that the wave-form of the emitted field is simply related to the shape of the pump pulse and, thereby, is easily controlled. The second observation to note is that the flux of cavity-mode photons tends to zero, as $\langle \sigma_{F}(t)\rangle \rightarrow 1$, i.e. when the atomic population is completely transferred into the state of $m_F=F$. This could happen even at the leading edge of the pump pulse indicating that for deterministic production of multi-photon FS no stringent restriction on laser
intensity is required (see Sec. III B).

\subsection{Production of the outgoing photons}

By integrating \eqref{nout15} using the new variable
\begin{equation}
\theta(t) = \int \limits_{-\infty}^t \alpha_1f_1(t')dt',
\end{equation}
we find for $n_{\text{out}}(t)$ a simple form
\begin{equation}
\label{noutfinal}
n_{\text{out}}(t) = \theta(t)-\int_0^{\theta(t)}\langle \sigma_{F}(\theta')\rangle d\theta'.
\end{equation}
This
shows that for small $\theta(t)$ the photon number increases proportionally to the pump energy confined in that area of the pulse.

The deterministic production of multi-photon FS requires
%no stringent restriction
that the laser intensity and cavity coupling be sufficiently strong such that (i) $\langle \sigma_F(t) \rangle \rightarrow 1$ (corresponding to large values of $\theta(t)$) and that (ii) the loss by spontaneous emission be negligible.  In these limits, $n_{\text{out}}$ is obviously an integer (see Sec. III A): $n_{\text{out}} = j$, that is a Fock state of $j$ photons is generated with $j$ in the range $1\leq j\leq 2F$, which depends on the state in which the atom has been initially prepared.
Condition (ii) is achieved when $\alpha_1f_1(t)\gg\Gamma_1(t)$.
This defines the signal-to-noise ratio
\begin{equation}\label{SN}
R_{\text{sn}} = {\alpha_1 f_1(t)\over \Gamma_{1}(t)} \simeq {4g^2\over k\gamma_{\text{sp}}},
\end{equation}
which needs to be large $R_{\text{sn}}\gg 1$ to achieve a high fidelity of deterministic photon production. This condition is in general expected to be fulfilled in high-finesse optical cavity with $g > k, \gamma_{\text{sp}}$.

Note that while being in the final state $m_F=F$, the atom continues to interact with the laser field, for example in the case of $^{87}$Rb atom, on the cycling transition $5S_{1/2}(F = 2, m_F=+2) \leftrightarrow 5P_{3/2}(F' = 3, m_{F'}=+3)$, without emitting a cavity photon, but experiencing only the Stark-shift of  ground state $m_F=2$.

%Another way to control the photon number is altering the optical transition of the atom,
%as is shown in the next section.

\section{\protect\normalsize EVOLUTION OF THE OUTPUT FIELD STATE. ANALYTIC AND NUMERICAL ANALYSIS}

In this Section, we investigate more specifically the production of a four-photon Fock state in the $5S_{1/2}(F = 2) \leftrightarrow 5P_{3/2}(F' = 3)$ hyperfine chain of $^{87}$Rb atom, which is prepared initially in the ground Zeeman state $|m_F=-F\rangle$ (see Fig. 1). We have optimized the cavity field and the laser Rabi frequency taking into account the model with the corresponding Clebsch-Gordan coefficients. In this case, the Rabi frequencies for the transitions $F,m_F\rightarrow F',m_{F'}=m_F+q$ by the $\sigma_+$ field (i.e. $q=1$) involve Wigner 3-$j$ and 6-$j$ symbols:
\begin{align}\label{OmCG}
\Omega_{m_{F},m_{F'}}&=(-1)^{m_F+J+I}\sqrt{(2F+1)(2F'+1)(2J+1)}\nonumber\\
&\;\times\left(\begin{array}{ccc} F & 1& F'\\ m_F &q &-m_{F'}\end{array}\right)
\left\{\begin{array}{ccc} J & J'& 1\\ F' &F &I\end{array}\right\}\Omega_1
\end{align}
with $J=1/2$, $J'=3/2$, $I=3/2$. The cavity field Rabi frequencies read (with the preceding coefficients for $q=0$)
\begin{align}\label{gCG}
g_{m_F,m_{F}}=&(-1)^{m_F+J+I}\sqrt{(2F+1)(2F'+1)(2J+1)}\nonumber\\
&\times\left(\begin{array}{ccc} F & 1& F'\\ m_F &0 &-m_{F}\end{array}\right)
\left\{\begin{array}{ccc} J & J'& 1\\ F' &F &I\end{array}\right\}g
\end{align}
with
\begin{equation}
g=\mu_{J,J'}\sqrt{\frac{\omega_c}{2\epsilon_0\hbar V}},
\end{equation}
where $\omega_c$ is the frequency of the cavity of volume $V$.
The partial spontaneous emission rates write ($q=0,\pm1$):
\begin{align}\label{gammaCG}
&\gamma_{F',m_{F}-q,F,m_F}=(2F+1)(2F'+1)(2J'+1)\nonumber\\
&\;\times\left[\left(\begin{array}{ccc} F & 1& F'\\ m_F &-q &-(m_{F}-q)\end{array}\right)
\left\{\begin{array}{ccc} J & J'& 1\\ F' &F &I\end{array}\right\}\right]^2\gamma_{\text{sp}}.
\end{align}

%It is worth noting
%that for off-resonant Raman process, as is the case here, $R_{\text{sn}}$ does not depend %on the detuning $\Delta$, which, however affects the efficiency of cavity-photon
%generation, as $\alpha_1 \sim \Delta^{-2}$. We suggest that for moderate pump
%intensities, $\Omega_1\sim g$, a detuning $\Delta \sim 10 g$ would be optimal.

We first derive closed-form formulas for the shape of the flux of photons and the produced number of photons when the spontaneous emission is neglected and assuming identical coupling constant for all transitions.

Numerical simulations are next shown for the complete model when spontaneous emissions is taken into account.

\subsection{Closed-formula in absence of spontaneous emission}

%To study the evolution of photonic states we will use the interaction picture, where an evolved
%state of the system can be found straightforwardly along with Hamiltonian \eqref{hamiltonian} as
%\begin{equation}\label{psi}
%|\Psi(t)\rangle = \exp\bigl ( -\int_{-\infty}^t H(t')dt')\bigr ) |\Psi(-\infty)\rangle,
%\end{equation}
%with initial state $|\Psi(-\infty)\rangle$ of the atom in an Zeeman state and the cavity mode in vacuum.
%However, instead of tedious and lengthy calculations, we find $|\Psi(t)\rangle$ in a simple way
%using the fact that the emission into spontaneous modes is greatly suppressed.

If one neglects all OP processes and assumes all dipole moments of optical transitions are the same, the basic states of the system are represented as a product of atomic Zeeman $|-F+j\rangle$ and photonic $|j\rangle$ states with $j$ the photon number. The state of the system $|\Psi(t)\rangle$ can be then expanded in this basis and yields
\begin{equation}\label{psi2}
|\Psi(t)\rangle = \sum_{j=0}^{2F}\beta_j(t) |-F+j\rangle|j\rangle,
\end{equation}
where the coefficients $|\beta_j(t)|^2 = \langle \sigma_{m_F=-F+j}\rangle$ are the populations of atomic states $|m_F=-F+j \rangle$ satisfying the normalization condition  $\sum_{j=0}^{2F}|\beta_j(t)|^2 = 1$ (see Eq. \eqref{sum}). They also describe the time variation of the photon distributions $P_j(t)=|\beta_j(t)|^2 $ in the resulting field. We solve Eqs. \eqref{mFock} with the initial
condition $\langle \sigma_{m_F=-2}(-\infty)\rangle =1$:
\begin{subequations} \label{solution}
\begin{eqnarray}
P_{0}(t)& =& e^{-\theta(t)},\label{p0} \\
P_{1}(t)& =& e^{-\theta(t)} \theta(t), \label{p1}\\
%P_{2} & =& e^{-\theta(t)} \int \limits_{-\infty}^t \alpha_1f_1(\tau)\theta(\tau)d\tau, \label{p2}\\
P_{2}(t) & =& e^{-\theta(t)} {\theta^2(t)\over 2}, \label{p2}\\
%P_{3} & =& e^{-\theta(t)} \int \limits_{-\infty}^t \alpha_1f_1(\tau)\int \limits_{-\infty}^{\tau}\alpha_1f_1(t')\theta(t')dt'd\tau, \label{p3}\\
P_{3}(t) & =& e^{-\theta(t)} {\theta^3(t)\over 6}, \label{p3}\\
%P_{4}& =& \int \limits_{-\infty}^t e^{-\alpha_1f_1(\tau)}\alpha_1f_1(\tau)\int \limits_{-\infty}^{\tau}\alpha_1f_1(t')\times \label{p4}\\
%& & \int \limits_{-\infty}^{t'}\alpha_1f_1(t'')\theta(t'')dt''dt'd\tau, \nonumber
P_{4}(t) & =& 1- e^{-\theta(t)} \biggl ( {\theta^3(t)\over 6} +{\theta^2(t)\over 2}+\theta(t)+1 \biggr ).
\end{eqnarray}
\end{subequations}
Closed-form formulas for the shape of the flux of photons and the produced number of photons are determined using \eqref{nout15} and \eqref{noutfinal}, respectively, with the obtained $P_4(t)$.

For a Gaussian laser pulse of intensity shape
\begin{equation}\label{gaus}
f_1(t)=e^{-(t/T)^2}
\end{equation}
of duration $T=1 \mu$s, the photon distributions are shown in Fig. 2, where we have taken the parameters
$(g, k,  \gamma_{\text{sp}}, \Omega_1, \Delta) = 2\pi\times(10, 3, 5.87, 10, 100)$ MHz, in order to fulfil all requirements, particularly, the condition $G_1 \ll k$. In this case, for the signal-to-noise ratio \eqref{SN} we have $R_{\text{sn}} \sim 20$, that justifies our approximations on neglecting OP processes.
It is seen that, as time increases, only the state with $j=4$ becomes populated in \eqref{psi2}, indicating that a four-photon FS is generated. In the intermediate region of time the output field is in a superposition of different photon-number states.

\begin{figure}[h] \rotatebox{0}{\includegraphics*
[scale =0.6]{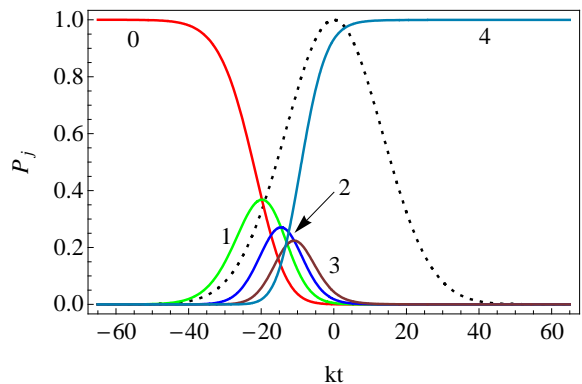}} \caption{(Color online) Dynamics of the photon number (indicated on the curves)
distribution of the generated light.
The atom is initially prepared in the state $m_F=-2$.
The Gaussian pulse shape \eqref{gaus} is shown by a dotted line.
\label{Fig2}}
\end{figure}

\subsection{Numerics}

Solving numerically Eqs. (4), taking into account the spontaneous emission and the actual coefficients [see Eqs. \eqref{OmCG}, \eqref{gCG}, and \eqref{gammaCG}], does not show any noticeable difference for the populations compared to the case without spontaneous emission. This is due to the fact that, in the considered system, spontaneous emission does not drive population outside the considered level-configuration.

\begin{figure}[h] \rotatebox{0}{\includegraphics*
[scale =0.7]{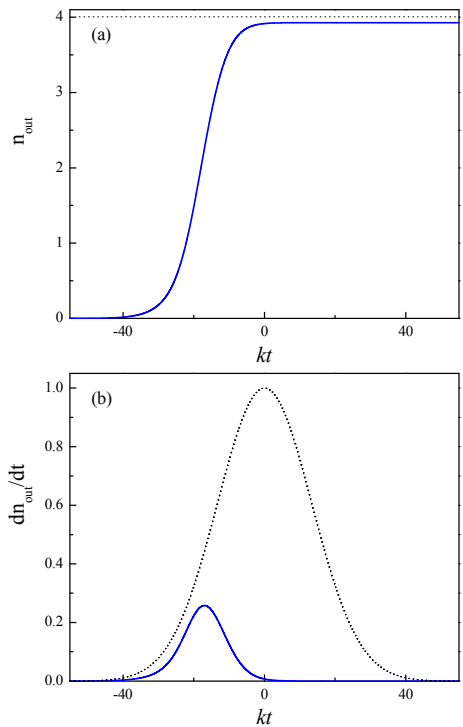}} \caption{(Color online) (a) Total number of output photons generated on the cycling $5S_{1/2}(F = 2) \rightarrow 5P_{3/2}(F' =
3)$ transition of $^{87}$Rb atom and (b) corresponding flux in the units of $k$. The Gaussian laser pulse
$f_1(t)$ is shown by dotted line. For the parameters, see the text.
\label{Fig3}}
\end{figure}

However, for the flux and the number of produced photons we find distinct results with or without spontaneous emission. This can be observed from Fig. 3 which displays the mean number $n_{\text{out}}$(t) of the output photons (a) and their flux (b). They are calculated numerically for the parameters $(g, k, \gamma_{\text{sp}}, \Omega_1, \Delta) = 2\pi\times(16, 1, 5.87, 10, 50)$ MHz, and $T = 3 \mu$s, by means of Eq. \eqref{nout} taking also into account all relaxation terms in Eqs. \eqref{mFock} and the actual coefficients from Eqs. \eqref{OmCG}, \eqref{gCG}, and \eqref{gammaCG} [in this case the condition $G_{1m_F, m_F+1}\ll k$ is fulfilled for all transitions owing to the Clebsch-Gordan coefficients in \eqref{OmCG}, \eqref{gCG}]. The total number of photons is indeed a bit less than the value $j=4$ expected in the ideal case discussed above. We find here $n_{\text{out}}(+\infty)\approx 3.93$. As we mentioned above, it is seen from Fig. \ref{Fig3} that the FS is generated at the leading edge of the laser pulse.

\subsection{Extension to a train of propagating Fock states}

Finally, we briefly discuss the possibility to produce a train of identical Fock-state-pulses on demand from our system that may have important quantum-information applications based on complex states of quantum light. Such deterministic and robust source of indistinguishable single photons has been proposed and analyzed in our previous paper \cite{SP}. Here we extend this technique to the case of multi-photon FS as follows. After the first $\sigma^+$ laser pulse has passed, a coherent $\sigma^-$ polarized laser pulse with the same frequency is turned on with a programmable delay time $\tau_d$, which is larger as compared to the pump pulse duration $T$ (see
Fig.1c). Then the process is running in the opposite direction, the atom is transferred by the $\sigma^-$ pump field from the state of $m_F=F$ to the state of $m_F=-F$ producing a Fock-state identical to the previous one. The pump repetition rate is not limited by any means that open wide possibilities for manipulation of the FS-stream up to constructing the superposition of FS. We remark that, in principle, using the parameters of Fig. 3, one should obtain $n_{\text{out}}(+\infty)\approx 3.93$ at each cycle. This means that the errors due to spontaneous emission do not accumulate since it does not drive atomic population outside the considered system. Thus, the implementation of this technique exploiting the cycling transitions of the atom is only limited by the atom lifetime in the cavity, which amounts to at most one minute \cite{12}.

\section{\protect\normalsize CONCLUSION}

In conclusion, we have proposed and analyzed a robust and realistic scheme to produce free-propagating Fock states on demand from a single atom trapped in a cavity QED and simultaneously interacting with a laser field system. The number of photons produced is predefined from the choice of the atomic hyperfine Zeeman structure and the initial condition of the $m_F$ state. In principle, from an hyperfine ground $F$-structure, one can produce a propagating Fock state comprising at maximum $2F$ photons. We have shown in particular that almost four photons can be produced in the $5S_{1/2}(F = 2) \leftrightarrow 5P_{3/2}(F' = 3)$ chain of $^{87}$Rb atom, initially prepared in the ground Zeeman state $|m_F=-2\rangle$.
With the atom initially prepared in the states of $m_F=-1,0$ or $1$ the same procedure would allow one to generate the FS with $j=3,2$ and $1$ photons, respectively. Other FS can be produced depending on the chosen atomic system and the Zeeman hyperfine structure. For instance, using the cesium $D_2$ cycling transition $6S_{1/2}(F = 4) \rightarrow
6P_{3/2}(F' = 5)$ the eight photon FS can be generated, while on the $6S_{1/2}(F = 3) \rightarrow 6P_{3/2}(F' = 2)$ transition of the same atom the FS with five- and less photons can be produced. The last case requires, however, care since the near-lying hyperfine state $F' = 3$ of the upper level $6P_{3/2}$ turns out to contribute to spontaneous decay outside the considered level-configuration.

One can extend our proposal to produce a train of identical propagating Fock-state pulses using alternating $\sigma^+$ and $\sigma_-$ laser pulses.

We have shown that the main constraint is given by the atomic spontaneous emission which imposes a sufficiently strong cavity field coupling. It can, however, be experimentally implemented with the state-of-the-art optical cavities.

%\bigskip
\subsection*{Acknowledgments}%\nonumber

This research has been conducted in the scope of the International Associated Laboratory (CNRS-France $\&$ SCS-Armenia)  IRMAS. We acknowledge additional support from the European Union Seventh Framework Programme  Grant No. GA-205025-IPERA, from the Science Basic Foundation of the Government of the Republic of Armenia, and from the Conseil R\'egional de Bourgogne.

\end{document}